\documentclass[amsmath,amssymb,preprint,showpacs]{revtex4}
\usepackage{graphicx}
\begin{document}

\title{Negative phase velocity in nonlinear oscillatory systems\\
---mechanism and parameter distributions}

\author{Zhoujian Cao$^{1}$, Pengfei Li$^{1}$, Hong Zhang$^{2}$, and Gang
Hu$^{1,3}$\footnote{Author for correspondence: ganghu@bnu.edu.cn}}

\address{$^{1}$Department of physics, Beijing Normal
University, Beijing,100875, China\\
$^{2}$Zhejiang Institute of Modern Physics and
Department of Physics, Zhejiang\\
University, Hangzhou 310027, China\\
$^{3}$Beijing-Hong Kong-Singapore Joint Center of Nonlinear and
Complex Systems, Beijing Normal University Branch, Beijing, China}

\begin{abstract}
Waves propagating inwardly to the wave source are called antiwaves
which have negative phase velocity. In this paper the phenomenon
of negative phase velocity in oscillatory systems is studied on
the basis of periodically paced complex Ginzbug-Laundau equation
(CGLE). We figure out a clear physical picture on the negative
phase velocity of these pacing induced waves. This picture tells
us that the competition between the frequency $\omega_{out}$ of
the pacing induced waves with the natural frequency $\omega_{0}$
of the oscillatory medium is the key point responsible for the
emergence of negative phase velocity and the corresponding
antiwaves. $\omega_{out}\omega_{0}>0$ and
$|\omega_{out}|<|\omega_{0}|$ are the criterions for the waves
with negative phase velocity. This criterion is general for one
and high dimensional CGLE and for general oscillatory models. Our
understanding of antiwaves predicts that no antispirals and waves
with negative phase velocity can be observed in excitable media.
\end{abstract}

\keywords{Oscillatory medium; CGLE; Antiwaves}

\maketitle

\section{Introduction}

A kind of novel waves, waves propagating inwardly to the
corresponding wave sources, were predicted theoretically in
sixties of the last century\cite{veselago}, and were
experimentally discovered at the beginning of this
century\cite{smith,vanag1}. Throughout this paper, we call this
special kind of waves antiwaves (AW) for short, which have
negative phase velocity, and call waves propagating forwardly from
the wave source normal waves (NW). The phenomenon of negative
phase velocity exists in both linear\cite{smith} and
nonlinear\cite{vanag1} systems. So far, linear optical antiwaves
have attracted great attention and have been introduced to applied
sciences\cite{lhm,topten1,aydin,smith2,pendry,topten2,grigorenko}.
While nonlinear antiwaves have been much less. Till now,
scientists have found only antispirals in nonlinear
systems\cite{vanag2,yang,gong,bar,kim}. In contrast to normal
spiral waves, antispiral waves propagate inwardly to the spiral
tips, which serve as the sources of the waves. In some recent
works, several models have been suggested to understand
antispirals\cite{vanag1,vanag2}. Gong and Christini have put
forward an empirical criterion for the emergence of
antispirals\cite{gong}. B\"{a}r \emph{et. al.} have explained
antispirals from the changes of the sign of $\omega$ (spiral
frequency) and $k$ (spiral wave number) with the model of
CGLE\cite{hagan}. While these studies illuminated some aspects of
antispiral waves, the mechanism and the general physical picture
of antiwaves are still not completely clear. With antispiral waves
in mind, some questions emerge naturally: Are there general
antiwaves exist (beside antispirals) which propagate toward the
wave sources? If yes, is there any common mechanism underlying
these antiwaves?

This paper will deal with the above problems. The paper is
organized as follows. In Section 2, we specify our nonlinear model
to produce travelling nonlinear AWs by certain local periodic
pacing. In Section 3, we analyze the mechanism of general
nonlinear AWs, including antispirals and planar travelling
antiwaves as their special cases. In Section 4, we specify the
parameter regions for emergence of AWs in CGLE. The CGLE system
serve as a simple example showing applications of our
understanding on nonlinear AWs. Since our analysis does not depend
on the detail dynamics of CGLE, we expect that the analysis in
this paper is generally valid for oscillatory media. We conclude
our results and give some remarks in Section 5.

\section{Model and nonlinear waves with negative phase velocity}

Oscillatory and excitable media are two kinds of media extensively
investigated in nonlinear dynamics. But antispirals are found only
in oscillatory media\cite{gong}. CGLE is a typical and well known
oscillatory system. Here we use 1D CGLE with local
pacing\cite{zhang} to explore nonlinear antiwaves.
\begin{eqnarray}
\frac{\partial A}{\partial
t}=A&-&(1+i\alpha)|A|^{2}A+(1+i\beta)\nabla^{2}A+\delta(\Omega)Fe^{-i\omega_{in}
t}\\
&&\delta(\Omega)=\left\{\begin{array}{l@{\quad}l}1&x\in\Omega\\
0&\mbox{otherwise}\\
\end{array}\right.\nonumber
\end{eqnarray}
We conduct simulations in an aray of $N (N=256)$ sites with
$\Delta x=1.0$ space discretization. Free boundary condition is
used throughout the paper. Numerically, we add the periodic signal
to the left boundary site of the 1D system. We fix $F=1.0$ for all
simulations of this paper. The sign of the input frequency
$\omega_{in}$ in Eq.(1) denotes the rotating direction of the
forcing in complex plane, ``+" clockwise and ``-" anticlockwise.
System (1) supports periodic travelling waves
\renewcommand{\theequation}{2\alph{equation}}
\setcounter{equation}{0}
\begin{eqnarray}
A&=&\sqrt{1-k^{2}}e^{i(-\omega t+kx)}\\
\omega&=&\alpha+(\beta-\alpha)k^{2}
\end{eqnarray}
Equation (2b) together with the condition $0\leq k^{2}\leq 1$
define the constraint of $\omega$,
\begin{eqnarray}
\alpha\leq \omega \leq \beta \mbox{ or } \beta\leq \omega \leq
\alpha
\end{eqnarray}

\begin{figure}
\includegraphics[ height = 4.0in, width = 0.6\linewidth]{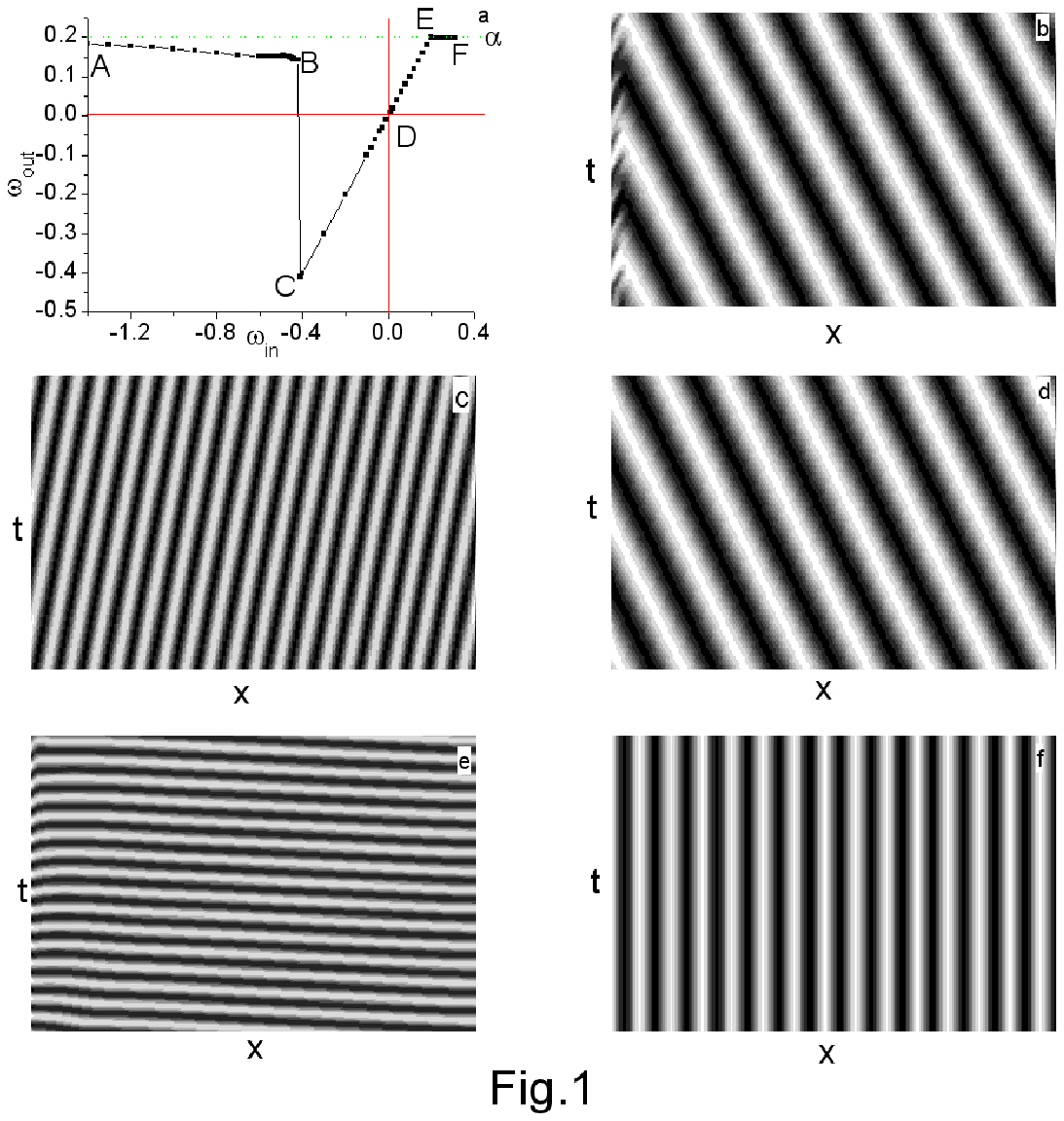}
\caption{Numerical results of Eq.(1). $\alpha=0.2, \beta=-1.4,
F=1$. (a) Input-output frequency relation. In AB and EF region,
constraint Eq.(2b) forbids the complete drift
$\omega_{out}=\omega=\omega_{in}$, and the system motion is
governed by the local dynamics, $\omega_{out}\approx\alpha$ and
$|k|\ll1$. CE region corresponds 1:1 resonant output
$\omega_{out}=\omega_{in}$. (b)-(f) Spatiotemporal patterns of the
system for different parameter sets. (b) $\omega_{in}=-0.5$ (AB
region), antiwaves. (c) $\omega_{in}=-0.15$ (CD region), normal
waves. (d) $\omega_{in}=0.15$ (DE region), antiwaves. (e)
$\omega_{in}=0.21$ (EF region). $\omega_{out}\approx\alpha=0.2$
and $k\approx0$, practically homogeneous oscillation. (f)
$\omega_{out}=\omega_{in}=0$, stationary Turning pattern.}
\end{figure}

In Fig.1(a) we plot the input-output frequency
relation\cite{stamp} for $\beta=-1.4, \alpha=0.2$. There are four
parameter domains in Fig.1(a), corresponding to different response
behaviors of the system to the local pacing. When
$\omega_{in}<-0.41$, the system cannot follow the pacing frequency
due to too fast pacing frequency, the output frequency of the
system is thus much lower than $\omega_{in}$ and also lower than
the natural frequency $(\omega_{0}=\alpha)$,
$|\omega_{out}|<|\omega_{0}|$. We observe antiwaves asymptotically
(see segment AB in Fig.1(a) and the evolution pattern in $t-x$
plane of Fig.1(b)). When $-0.41<\omega_{in}<0$, the system
oscillates with the input frequency $\omega_{out}=\omega_{in}$
resonantly and the waves propagate normally (CD segment in
Fig.1(a) and Fig.1(c)). When $0<\omega_{in}<0.2$, the system
oscillates also with the input frequency $\omega_{in}$ while waves
propagate inwardly to the wave source, i.e., one observes negative
phase velocity (DE segment in Fig.1(a) and Fig.1(d)). When
$\omega_{in}>0.2$, the system approaches asymptotically to nearly
homogeneous oscillation, i.e. $k\ll 1$,
$\omega_{out}\lesssim\omega_{0}=\alpha$ (antiwaves with very large
phase velocity, EF segment in Fig. 1(a) and Fig.1(e)). An
interesting point in Fig.1 is that we can produce both NWs and AWs
with the same parameter set of $(\alpha,\beta)$ by changing the
pacing frequency only. There is a special case of $\omega_{in}=0$,
which locates at the boundary between NWs and AWs. We expect that
this local constant force produces neither normal waves nor
antiwaves. Indeed we observe in Fig.1(f) a pacing induced
stationary Turning pattern.

The above numerical results demonstrate that local pacing with
suitable frequency can produce AWs in nonlinear oscillatory
systems indeed. These results show that nonlinear NWs and AWs are
determined by the competition between the output frequency
$\omega_{out}$ and the local natural frequency $\omega_{0}$
($\omega_{0}=\alpha$ for CGLE) of oscillatory media. In summary,
if $\omega_{out}$ and $\omega_{0}$ have the same sign while the
absolute value of $\omega_{out}$ is smaller than that of
$\omega_{0}$, we have
\renewcommand{\theequation}{3\alph{equation}}
\setcounter{equation}{0}
\begin{eqnarray} \omega_{out}\omega_{0}>0\mbox{
and}|\omega_{0}|>|\omega_{out}|\mbox{ \ \ \ Antiwaves}
\end{eqnarray}
In other two cases we have
\begin{eqnarray}
\omega_{out}\omega_{0}>0\mbox{ and
}|\omega_{0}|<|\omega_{out}|&&\mbox{ \ \ \ Normal waves}\\
\omega_{out}\omega_{0}<0&&\mbox{ \ \ \  Normal waves}
\end{eqnarray}
It is thus the competition between $\omega_{out}$ and $\omega_{0}$
that makes the system to select the correct sign of $k$ in the
dispersion relation Eq.(2b), and to determine positive
($k\omega>0$) or negative ($k\omega<0$) phase velocity (or, say,
normal or anti waves).

It is emphasized that in Eq.(3) we use frequency relation between
$\omega_{out}$ and $\omega_{0}$, not $\omega_{in}$ and
$\omega_{0}$. Here $\omega_{out}$ is the actual frequency of the
system motion. In certain cases, the oscillatory system is
completely driven by the pacing and we have
$\omega_{out}=\omega_{in}$. Then $\omega_{in}$ can be used instead
of $\omega_{out}$ for classifying the cases of Eqs.3(a), (b) and
(c) (see regions CD and DE in Fig.1(a)). In some other cases, the
system cannot follow the rotation frequency of the pacing, and the
asymptotic state of the system has frequency
$\omega_{out}\neq\omega_{in}$, then all the above analysis
describes how the competition between $\omega_{out}$ and
$\omega_{0}$ produces normal and anti waves in oscillatory media.
This is what happens in the segments AB and EF of Fig.1(a).

\section{Mechanism underlying nonlinear antiwaves}

Now we try to understand the phenomena of negative phase velocity
and antiwaves in CGLE. Firstly we begin with the analysis of the
local dynamics of the system. Without the coupling and forcing
terms the local dynamics of CGLE reads \setcounter{equation}{3}
\renewcommand{\theequation}{\arabic{equation}}
\begin{eqnarray}
\frac{d A}{d t}=A-(1+i\alpha)|A|^{2}A
\end{eqnarray}
Representing complex variable $A$ by $A=Re^{i\theta}$, we can
transform Eq.(4) to
\begin{eqnarray}
\left\{\begin{array}{l@{\quad}l}\frac{d R}{d t}=R(1-R^{3})\\
\frac{d \theta}{d t}=-\alpha R^{2}\\
\end{array}\right.
\end{eqnarray}
Eq.(5) is a typical limit cycle system with frequency $\alpha$.
Every grid of system (1) is a rotator $e^{-i\alpha t}$ who rotates
clockwise (anticlockwise) in complex plane if $\alpha$ is positive
(negative). Therefore, Eq.(1) is an oscillatory medium, and we
call $\alpha$ the local natural frequency of the system, denoted
as $\omega_{0}=\alpha$.

When we force the left boundary of the system, the grids near the
left boundary are driven by the pacing, and they can quickly reach
their asymptotic rotation with the output frequency
$\omega_{out}$. Due to the diffusion coupling, the perturbation
from the left boundary can stimulate the grids distance away from
the left boundary, and make these grids rotate with their natural
frequency $\omega_{0}$ in the first evolution stage. In the next
evolution stage, the grids away from the pacing source will be
synchronized to the asymptotic frequency $\omega_{out}$, because
the coupling restricts the phase difference between these grids
and the grids near the left boundary. It is just the initial phase
arrangement of different grids by $\omega_{out}$ and $\omega_{0}$
plays the key role in determining the sign of the phase velocity.
\begin{figure}[bt]
\includegraphics[ height = 4.0in, width = 0.6\linewidth]{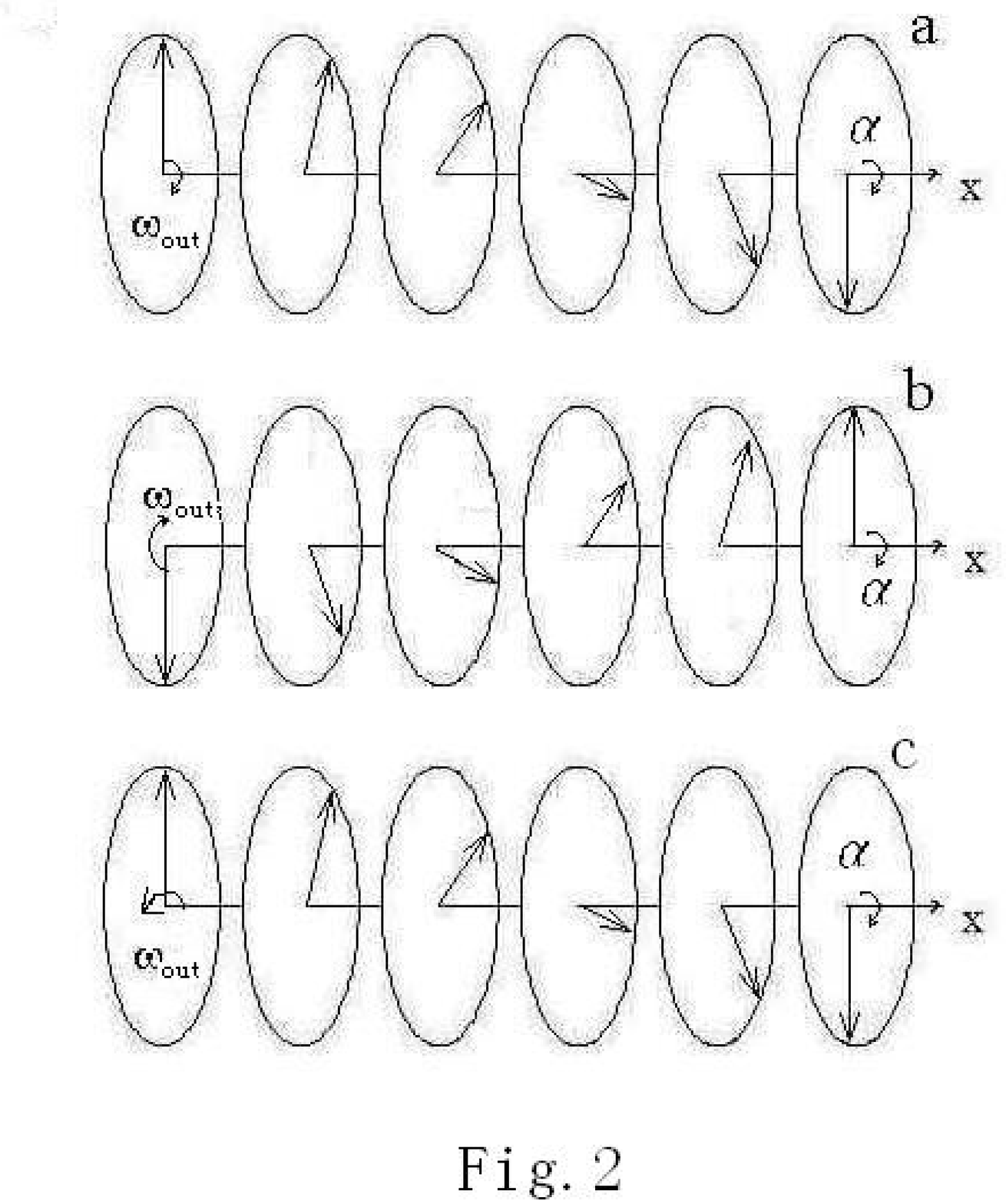}
\caption{Spatial phase distribution responding the perturbation
from the wave source. Frequency $\omega_{0}$ is supposed in
clockwise direction. (a) $\omega_{out}\omega_{0}>0$ and
$|\omega_{out}|>|\omega_{0}|$. Given a space site (with frequency
$\omega_{out}$), the sites farther away from the pacing, having
frequencies (near $\omega_{0}$) larger than $\omega_{out}$, rotate
faster than the given site. When these sites are synchronized to
the frequency $\omega_{out}$, they must have phases more advanced
than the phase of the given site. This phase distribution results
in negative phase velocity and AWs. (b)
$\omega_{out}\omega_{0}>0$, $|\omega_{out}|>|\omega_{0}|$ and (c)
$\omega_{out}\omega_{0}<0$. By the same argument as (a), the sites
farther away from the pacing source have phase delay with respect
to the given site. We can observe positive phase velocity and
NWs.}
\end{figure}

Now we analyze the three types of Eq.(3). First, we suppose that
the pacing generated frequency $\omega_{out}$ is in the same
direction with the local natural frequency $\omega_{0}$ and we
have $|\omega_{0}|>|\omega_{out}|$ (case Eq. (3a)). The grids near
the left boundary rotate initially slower than its neighbor grids
farther away from the pacing source. When synchronization of all
these grids are reached the phase of the latter must be more in
advance than that of the former. One thus observes phase
propagation towards to the pacing source, an interesting
phenomenon of negative phase velocity. This kind of phase setting
is shown in Fig.2(a).

Second, if $\omega_{out}$ and $\omega_{0}$ have the same sign
while $|\omega|>|\omega_{0}|$ (case Eq.(3b)), the initial phase
distribution of the grids is such prepared for the asymptotical
travelling waves that any grid in the 1D chain must have some
phase delay in comparison with its neighbor grids nearer the
pacing source as shown in Fig.2(b). In this case one can observe
normal forward propagating waves. Third, if the two frequencies
$\omega_{out}$ and $\omega_{0}$ have opposite rotation directions
($\omega_{0}\omega_{out}<0$, case Eq.(3c)), one can observes phase
distribution of grids of Fig.2(c) yielding normal forward waves.

Now we can give some clear conclusions on the problem of antiwaves
and negative phase velocity. These conclusions are valid not only
for CGLE systems, but also for general oscillatory systems. We
suppose that a system is oscillatory with local dynamics having
bulk frequency $\omega_{0}$ \cite{vanag1,gong} and the medium
supports periodic waves with frequency $\omega_{out}$ generated by
an arbitrary perturbation source. If $\omega_{out}$ and
$\omega_{0}$ have the same sign and $|\omega_{out}|<|\omega_{0}|$,
we can surely produce antiwaves, and we can produce normal waves
in all other cases, except two no generic critical cases: $k=0$
(producing homogeneous oscillations) and $\omega_{out}=0$
(producing stationary Turning patterns). It is emphasized that our
analysis based on the pictures of Fig.2 does not use any special
dynamic structure of CGLE, and thus does not need the requirement
that the system is near Hopf bifurcation condition. This
generality has never been explored in previous analysis on the
problems of negative phase velocity. In our analysis, local
oscillation (i.e., nonzero $\omega_{0}$) is a necessary condition
for negative phase velocity. If the local dynamics of a medium
does not have oscillation, we have $\omega_{0}=0$. It is clear
from Eq.(3) that the system with $\omega_{0}=0$ can support only
normally propagating waves. All excitable media have local
frequency $\omega_{0}=0$, and we can therefore answer a previously
raised question about antispirals in excitable media\cite{gong}:
antiwaves (including antispirals) can never be found in excitable
media. We have numerically computed CGLE for many other parameter
sets, and computed different chemical reaction diffusion models
with oscillatory and excitable local dynamics, the above
conclusions have been fully confirmed.

\section{Parameter domain for antiwaves in CGLE
system}

The analysis of Eq.(3) and the physical mechanism of Fig.2 are
valid for general oscillatory systems. CGLE is a kind of
oscillatory system universally appearing around Hopf bifurcation
from a homogeneous stationary states to homogeneous oscillation.
It is thus interesting to classify parameter domains for normal
and anti waves of CGLE. In this section we focus on specifying the
domain of antiwaves of CGLE in $(\alpha,\beta,\omega_{out})$
parameter space.

The motion of periodically forced CGLE can be periodic,
quasiperiodic and chaotic. Here we are interested only in periodic
travelling wave states. Far away from the pacing boundary,
travelling waves must have solution of Eq.(2a) and the frequency
$\omega_{out}$ must obey condition Eq.(2c). This condition is
required for all periodic travelling waves, including both NW and
AW. On the other hand, the conditions of Eq.(3) can be used to
classify NW and AW. Eqs.(2c) and (3) together completely determine
the parameter domains of NW and AW of CGLE.
\begin{figure}[bt]
\includegraphics[ height = 4.0in, width = 0.6\linewidth]{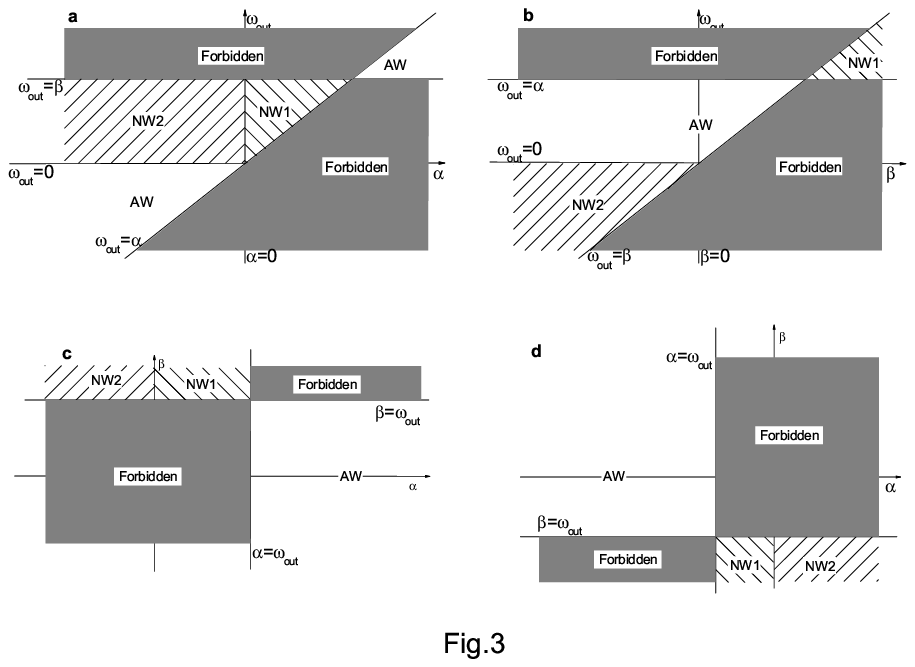}
\caption{Schematic figures of phase diagrams of CGLE, which are
drawn, based on Eq.(2c) and (3). (a) Phase diagram in
$\omega_{out}-\alpha$ plane with $\beta>0$ fixed. In the grey
regions marked ``Forbidden", travelling waves cannot be realized
due to the condition of Eq.(2c). The blank region marked AW is
antiwave region, characterized by Eq.(3a). The shadowed region
marked NW1 (NW2) is a part of normal wave region, characterized by
Eq.(3b) (by Eq.(3c)). (b) The same as (a) with $\alpha>0$ fixed
and the distribution of various characteristic domains is plotted
in $\beta-\omega_{out}$ plane. (c) The same as (a) with
$\omega_{out}>0$ fixed and the domain distribution is plotted in
$\alpha-\beta$ plane. (d) The same as (c) with $\omega_{out}<0$.
(d) and (c) are symmetric against the symmetry center
$(\alpha,\beta)=(0,0)$.}
\end{figure}

In Fig.3(a) we present domain distribution of NWs and AWs in
$\alpha-\omega_{out}$ plane with $\beta$ fixed. In the parameter
region marked ``Forbidden", the parameters do not satisfy the
constraint of Eq.(2c), and no periodic travelling wave exists in
this domain. In the blank region (the blow-left and up-right
parts), the parameters satisfy Eq.(2c) and the conditions of
Eq.(3a), we can observe AWs. In the shadowed region marked NW1
(NW2), the parameters satisfy Eq.(2c) and Eq.(3b) (and Eq.(3c)),
and NWs are produced. In Figs.3(b) and 3(c) we do exactly the same
as Fig.3(a) with $\alpha$ and $\omega$ fixed, respectively. Figs.
3(a), 3(b) and 3(c) are obtained for arbitrary positive $\beta$,
$\alpha$ and $\omega_{out}$. For negative $\beta$, $\alpha$ and
$\omega_{out}$ we can directly obtain domain distribution diagrams
from Fig.3(a), 3(b) and 3(c), replacing $\alpha$, $\omega_{out}$;
$\beta$, $\omega_{out}$; and $\alpha$, $\beta$ by $-\alpha$,
$-\omega_{out}$; $-\beta$, $-\omega_{out}$; and $-\alpha$,
$-\beta$, respectively (e.g., see Fig.3(d)).
\section{Conclusion}
In conclusion, we have found interesting phenomenon of negative
phase velocity (antiwaves) induced by local pacing, and revealed
the common mechanism underlying antiwaves: the competition between
the actual system frequency $\omega_{out}$ and the local natural
frequency $\omega_{0}$ selects the correct sign of the wave number
$k$ in the dispersion relation Eq.(2b) to produce either normal
waves or antiwaves. If condition $\omega\omega_{0}>0$ and
$|\omega|<|\omega_{0}|$ are satisfied, antiwaves can emerge
definitely. To our best knowledge, it is the first time to
emphasize that both signs and values of $\omega$ and $\omega_{0}$
play essential roles in producing antiwaves. This understanding
gives a convincing answer to the problem why no antispirals and
antiwaves have been observed in excitable media. The criterion of
antiwaves (Eq.(3a) and Fig.2(a)) is expected to be valid for
general oscillatory systems, and this goes again beyond the
previous results of negative phase velocity of CGLE.

\end{document}